\documentclass[twocolumn]{aa}

\usepackage{natbib}
\usepackage{graphicx}
\usepackage{amsmath,amssymb,bm}
\usepackage{threeparttable,multirow,makecell}
\usepackage[figuresright]{rotating}

\usepackage[colorlinks=true,
            linkcolor=black,
            citecolor=black,
            urlcolor=blue]{hyperref}

\usepackage{balance}

\makeatletter
\def\linenumbers{}
\makeatother

\titlerunning{Background dynamics and constraints of $\Lambda(t)$CDM models}
\authorrunning{O.Avsajanishvili}

\begin{document}

\title{Background dynamics and observational constraints of spatially flat and non-flat $\Lambda(t)$CDM models from H(z) and DESI DR2 BAO measurements}  

\author{Olga Avsajanishvili\inst{1}}  

\institute{ 
Evgeni Kharadze Georgian National Astrophysical Observatory, 0179 Tbilisi, Georgia\\
\email{olga.avsad@gmail.com}
}  

\date{Received XXX; accepted XXX}  

\abstract
{Dynamical vacuum models provide a minimal phenomenological extension of the standard $\Lambda$ cold dark matter ($\Lambda$CDM) model in which the vacuum energy density evolves with cosmic time, potentially affecting late-time expansion.}
{We investigated the background evolution of spatially flat and non-flat $\Lambda(t)$CDM cosmologies, parametrised by a dimensionless vacuum parameter, $\alpha$, through their effective equation of state (EoS) and deceleration parameter, as well as their expansion dynamics. We aim to constrain their parameters using observational data and to assess their performance relative to spatially flat and non-flat $\Lambda$CDM models.}
{We derived analytical expressions for the total effective and dark energy EoS parameters, as well as the deceleration parameter. We performed parameter estimation using Markov chain Monte Carlo (MCMC) analysis with 32 cosmic chronometer $H(z)$ measurements and 13 DESI DR2 baryon acoustic 
oscillation (BAO) data points. Model comparison was carried out using $\chi^2_{\min}$, Akaike, corrected  Akaike, Bayesian, and deviance information criteria, and Akaike weights.}
{The $\Lambda(t)$CDM model reproduces the standard matter-dominated evolution at early times and approaches a de Sitter phase at late times. The vacuum parameter, $\alpha$, modifies the expansion history, shifting the transition from matter domination to dark energy domination, and altering the timing of the acceleration transition. Constraints from $H(z)$ data alone are weak and exhibit parameter degeneracies, particularly in the spatially non-flat case. The inclusion of DESI DR2 BAO data significantly tightens constraints and yields $\alpha$ consistent with zero within $1\sigma$ in both spatially flat and non-flat scenarios. The spatial curvature is constrained to values consistent with flatness. The inclusion of BAO data also reduces the Hubble tension with early Universe measurements to below the $1\sigma$ level, while a residual $\sim 2.6$–$3.2\sigma$ tension with late-time determinations remains. Information criteria favour the spatially flat $\Lambda$CDM model.}
{Current expansion data show no statistically significant evidence of vacuum dynamics. The $\Lambda(t)$CDM model is observationally viable but is not preferred to the standard spatially flat $\Lambda$CDM cosmology.}


\maketitle  

\section{Introduction}\label{section:1}
The standard spatially flat cosmological Lambda cold dark matter ($\Lambda$CDM) model~\citep{1984ApJ...284..439P,Peebles:2002gy,Lopez-Corredoira:2023jhh} has demonstrated good consistency with a wide range of observational probes accumulated over the last several decades~\citep{Planck:2018nkj,Planck:2018vyg,ACT:2020gnv,eBOSS:2020yzd}. Nevertheless, despite its status as the current concordance model, the $\Lambda$CDM model exhibits a number of statistically significant tensions and anomalies~\citep{Perivolaropoulos:2021jda,DiValentino:2022fjm,Abdalla:2022yfr,Peebles:2022akh,Vagnozzi:2023nrq}.
One of the most prominent of the various discrepancies reported in the literature is the tension in the Hubble constant, $H_0$, manifested as a statistically significant disagreement between the value inferred from $Planck$ cosmic microwave background (CMB) observations~\citep{Planck:2018vyg} and that obtained from local distance-ladder measurements by the Supernova $H_0$ for the Equation of State (SH0ES) team~\citep{Riess:2021jrx}. 
In this context, another long-standing open issue in modern cosmology is the so-called coincidence problem, namely the question of why the energy densities of dark energy (DE) and dark matter (DM) are of the same order of magnitude precisely at the present cosmic epoch~\citep{Steinhardt:1999nw, Garriga:2000cv, Dalal:2001dt, Ishak:2005xp, Starkman:2006at, Maor:2007ygp, Martin:2012bt, Zheng:2021uee}. 

Several approaches have been proposed to address this problem. An earlier and widely studied one is based on attractor (tracker) solutions in self-interacting scalar field dynamical DE models~\citep{Ratra:1987rm, Ratra:1987aj}. In addition, interacting DE (IDE) models~\citep{Wetterich:1994bg, Amendola:1999dr, Amendola:1999er, Zimdahl:2001ar, Liu:2023mwx} provide an alternative framework by allowing energy exchange between DE and DM, thereby enabling a dynamical interplay between these two components. 
Such interactions can naturally modify the late-time expansion history of the Universe, providing a viable mechanism to alleviate the coincidence problem and, through the resulting changes in the background dynamics, offering a phenomenological avenue to ease the observed $H_0$ tension~\citep{Yang:2018euj, DiValentino:2019ffd, DiValentino:2019jae, Nunes:2021zzi, DiValentino:2021izs, Nunes:2022bhn, Bernui:2023byc, Hoerning:2023hks, Teixeira:2023zjt,  Zhai:2023yny, Pan:2023mie, Blanchard:2025jff}.

The time-varying Lambda cold dark matter ($\Lambda(t)$CDM) model~\citep{Benetti:2019lxu,Benetti:2021div, DiValentino:2021izs, Borges:2023xwx, Koussour:2024nhw, Benetti:2024dob, Yang:2025vnm, diDonato:2025tbd, Guillen:2025hix} belongs to a wide class of IDE scenarios. This model represents a phenomenological extension of the standard
$\Lambda$CDM cosmology in which the vacuum energy density is allowed to evolve with
cosmic time and interact with the matter sector. 
The interaction between matter and vacuum energy is parametrised by a single dimensionless vacuum dynamics parameter, $\alpha$\footnote{The $\Lambda(t)$CDM model is closely related to the running vacuum model (RVM)~\citep{Shapiro:2003ui,Sola:2016hnq,SolaPeracaula:2018wwm,SolaPeracaula:2022hpd, Moreno-Pulido:2022phq, Moreno-Pulido:2022up,SolaPeracaula:2023swx,Moreno-Pulido:2023ryo, SolaPeracaula:2026trz}, as both scenarios belong to the class of cosmologies with dynamical vacuum energy interacting with matter. While RVM is typically formulated through an explicit dependence of the vacuum energy density on the Hubble parameter, $\rho_\Lambda=\rho_\Lambda(H)$, motivated by quantum field theory in curved space-time, the $\Lambda(t)$CDM framework implements vacuum dynamics phenomenologically via a modified matter dilution law. At the background level and for small values of the model parameter, $\alpha$, the two models exhibit nearly equivalent expansion histories, leading to similar late-time cosmological signatures.}, which quantifies departures from the standard $\Lambda$CDM model and ensures a continuous $\Lambda$CDM limit for $\alpha=0$, which is fully recovered only asymptotically in the future.

At the background level, this model can be interpreted as equivalent to a generalized Chaplygin gas, providing a phenomenological unified description of DM and DE  as a single fluid.  The concept of unifying DM and DE was first introduced by Chaplygin~\citep{Chaplygin:1904}, and the original Chaplygin model, along with its various generalisations, has been extensively studied both theoretically and observationally~\citep{Bilic:2001cg, Caldwell:2005xb,Benaoum:2002zs,Avelino:2003cf,Debnath:2004cd,Guo:2005qy,Banerjee:2005vy,delCampo:2008vr, Lu:2008hp,2008PhLB..662...87L,2009EPJC...63..349L,2011GReGr..43..819L,2010JCAP...03..025X,2011A&A...527A..11L,Bouhmadi-Lopez:2011tfh,2012EPJC...72.1931X,vomMarttens:2017cuz}. The generalized Chaplygin gas arises from a barotropic fluid with the equation of state (EoS) $p = -A/\rho^\alpha$,
  where $p$  and $\rho$ are the pressure and the energy density of the barotropic fluid, respectively; $ A > 0 $ is a constant and $\alpha$ is the model parameter (typically $\alpha > -1$ to avoid singularities)\footnote{For the original Chaplygin gas, $\alpha=1$ and the EoS has the form $p = -A/\rho$.}. It was originally proposed to account for the observed cosmic acceleration without invoking separate DM and DE components~\citep{Kamenshchik:2001cp,Bento:2002ps,Gorini:2002kf,Amendola:2003bz,Multamaki:2003ed,Gorini:2005nw,Banerjee:2006na,Lu:2008hp,2012EPJC...72.1883X,Lu:2013una,Herrera:2016sov}. In this framework, the unified fluid behaves like DM during the early stages of cosmic evolution  and asymptotically approaches a cosmological constant at late times, with vacuum energy interacting with matter. 

In this work, we reconstructed the total effective EoS parameter of the unified cosmic fluid, the DE EoS, and the deceleration parameters within the framework of the  spatially flat $\Lambda(t)$CDM model. We performed a detailed analysis of the dynamical behaviour of the model by examining the expansion history of the Universe, along with the temporal evolution of the total effective EoS parameter of the unified cosmic fluid and the deceleration parameter.

Direct measurements of the Hubble parameter from cosmic chronometers provide a relatively model-independent probe of the late-time expansion history. When combined with baryon acoustic oscillation (BAO) measurements, these data enable a robust assessment of vacuum dynamics and spatial curvature, significantly reducing the parameter degeneracies present when $H(z)$ data are used alone. We applied Markov chain Monte Carlo (MCMC) analysis to constrain the parameters of spatially flat and non-flat $\Lambda(t)$CDM models and compared them with the standard spatially flat and non-flat $\Lambda$CDM scenarios using $H(z)$ measurements~\citep{Simon:2004tf,Stern:2009ep,2014RAA....14.1221Z,Ratsimbazafy:2017vga,Borghi:2021rft,Cao:2023eja} and combined $H(z)$+BAO data from the Dark Energy Spectroscopic Instrument (DESI) Data Release 2 (DR2)~\citep{DESI:2025zgx}. To our knowledge, this represents the first observational analysis of the spatially flat and non-flat $\Lambda(t)$CDM models based on DESI DR2 BAO data. For model comparison, we employed several Bayesian statistical criteria and goodness-of-fit estimators, including $\Delta\chi^2_{\min}$, the Akaike information criterion (AIC), the corrected Akaike information criterion (AICc), the Bayesian information criterion (BIC), the deviance information criterion (DIC), and the Akaike weights $w_i$. Rather than achieving a full resolution of the Hubble tension, our goal is to assess whether vacuum dynamics within the $\Lambda(t)$CDM framework can alleviate the discrepancy between early- and late-time determinations of the Hubble constant, $H_0$.

This paper is organised as follows: Section~\ref{section:2} presents a description of the $\Lambda(t)$CDM model; Section~\ref{section:3} discusses the derivation of the reconstructive formulas for the total effective EoS parameter of the unified cosmic fluid, the DE EoS parameter, and the deceleration parameter in the spatially flat $\Lambda(t)$CDM model; Section~\ref{section:4} considers the features of observational constraints on the parameters of the spatially flat and non-flat $\Lambda$CDM and $\Lambda(t)$CDM models; Section~\ref{section:5} presents the results and analysis of the calculations; and Section~\ref{section:6} summarises the conclusions.
 We use the natural system of units here: $c=k_B=1$.

\section{Background dynamics of the $\Lambda(t)$CDM model}
 \label{section:2}

We assumed that the flat, homogeneous and isotropic Universe is described by the Friedmann–Lemaître–Robertson–Walker (FLRW) space-time metric $ds^2 = dt^2 -a^2(t)d{\bf x}^2$, where $t$ is a cosmic time; ${\bf x}$ are the comoving coordinates; $a(t)$ is the scale factor, normalised to be unity at the present epoch $a_0 \equiv a(t_0)=1$, so that $a = 1/(1 + z)$, here $z$ is the redshift.

Within the FLRW framework, we considered a Universe filled with pressureless matter composed of cold dark matter (CDM) and baryons, and a dynamical vacuum-like energy density $\Lambda(t)$ depending on cosmic time. We neglected the contribution of massive neutrinos, which is sub-dominant at the background level in the redshift range considered. We assumed that only the CDM sector interacts with the vacuum component, while baryons remain separately conserved and interact only gravitationally.

The first Friedmann equation (we adopted units such that $8\pi G/3 = 1$) reads
\begin{equation}
3H^2 = \rho_{cdm} + \rho_b + \Lambda(t),
\end{equation}
while the corresponding conservation equations are
\begin{eqnarray}
\dot{\rho}_{cdm} + 3H\rho_{cdm} = -\dot{\Lambda}(t) = \Gamma \rho_{cdm}, \\
\dot{\rho}_b + 3H\rho_b = 0,
\end{eqnarray}
where $\rho_{cdm}$ and $\rho_b$ are the energy densities of CDM and baryons, respectively; the over-dot denotes differentiation with respect to cosmic time, and $H$ is the Hubble parameter.

In this framework, the total matter energy density is given by $\rho_m = \rho_{cdm} + \rho_b$, while the vacuum interaction is assumed to act exclusively within the DM sector. This construction ensures consistency with local gravity constraints, in particular with experimental tests of the equivalence principle (Eötvös-type experiments)~\citep{Adelberger:2003zx, Will:2014kxa}. 

The dependence of the dynamical cosmological constant on the Hubble parameter is represented as~\citep{Benetti:2019lxu,Benetti:2021div} 
\begin{equation}\label{eq:Lambda}
\Lambda(t)=\varrho H^{-2\alpha},~{\rm with }~\varrho=3(1-\Omega_{m0})H_0^{2(1+\alpha)}. 
\end{equation}
Here  $H_0$ is the value of the Hubble parameter at the present epoch; $\Omega_{m0}$ denotes the total matter density parameter at the present epoch, including both baryonic and CDM components.
The constant $\alpha$ is the vacuum dynamics parameter (also acting as an interaction parameter), whose absolute value quantifies the strength of energy exchange between DE and matter over cosmic time. It parameterises deviations from the $\Lambda$CDM model and characterises the dynamical nature of the vacuum component. As mentioned above, this parameter is restricted to $\alpha > -1$. 
The function $\Gamma$ specifies the effective rate of matter creation for $\Gamma > 0$ and the annihilation rate for $\Gamma < 0$, and in this class of models takes the form of~\citep{Benetti:2019lxu,Benetti:2021div} 
\begin{equation}\label{eq:Gamma}
\Gamma = -\alpha \varrho H^{-(2\alpha+1)}.
\end{equation}
In the $\Lambda(t)$CDM model, the general form of the normalised Hubble parameter, in which the DE and matter components are unified and parameterised by a power-law dependence on the vacuum dynamics parameter,
$\alpha$, is given by~\citep{Benetti:2019lxu,Benetti:2021div} 

\begin{multline} 
E(a)=\bigg[\Big((1-\Omega_{m0})+\Omega_{m0}a^{-3(1+\alpha)}\Big)^{\frac{1}{(1+\alpha)}} + 
\\+\Omega_{r0}a^{-4} + \Omega_{k0}a^{-2}\bigg]^{1/2},
\label{eq:E} 
\end{multline} 
where $E = H(a)/H_0$ is the normalised Hubble parameter; $\Omega_{r0}$ and $\Omega_{k0}$ denote the radiation and spatial curvature density parameters at the present epoch, respectively. In this work, we adopted fiducial values $H_0 = 68.91~{\rm km~s^{-1}Mpc^{-1}}$ and $\Omega_{m0} = 0.3038$ from~\citet{DESI:2025hao} for the construction of background quantities\footnote{These values are used only for the construction of background and diagnostic quantities such as $H(a)$, $E(a)$, $q(a)$, and $w_{tot}(a)$. They are not fixed in the statistical analysis and are not included in the MCMC sampling.}.
For positive values of the vacuum dynamics parameter $\alpha$, the vacuum energy density decreases more rapidly than in the standard $\Lambda$CDM scenario, leading to a slower dilution of the matter component. Negative values of $\alpha$ produce the opposite behaviour: the vacuum energy density decays more slowly than in the standard $\Lambda$CDM model, while the matter component dilutes more rapidly.
In both cases, the parameter $\alpha$ modifies the relative evolution of the matter and vacuum sectors at the background level without violating the separate conservation of matter in the $\Lambda(t)$CDM framework. The limit $\alpha = 0$ recovers the standard $\Lambda$CDM model, in which matter and vacuum energy densities evolve independently.

\begin{figure}[!th]
\begin{center}
\includegraphics[width=1\columnwidth]{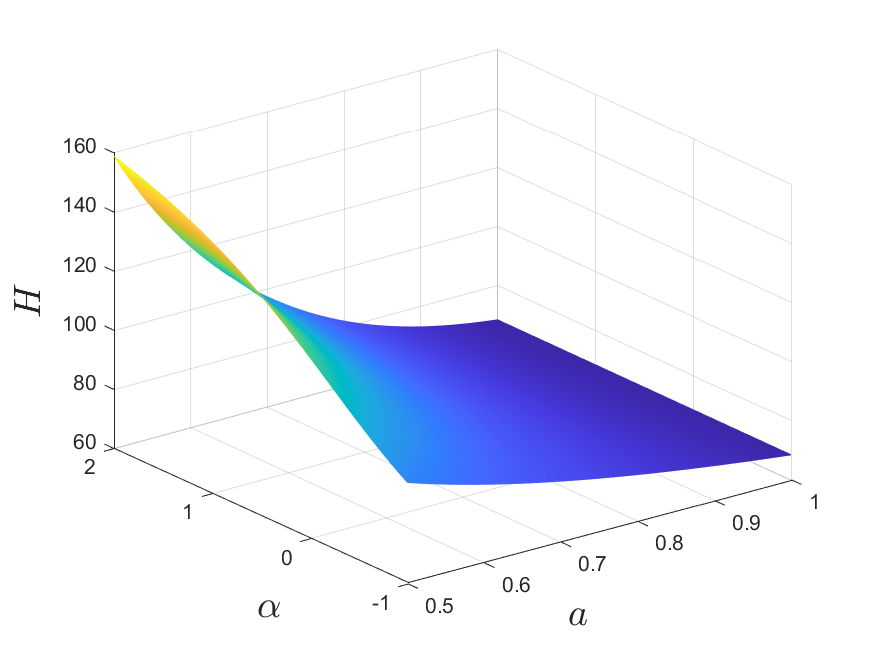}
\end{center}
\caption{Three-dimensional representation of the Hubble parameter, $H(a)$, in the spatially flat $\Lambda(t)$CDM model, shown in the $\alpha-a$ phase space.}
 \label{fig:f1}
 \end{figure}

By analysing Eq.~(\ref{eq:E}), it follows that in the early Universe ($a \to 0$), the term proportional to $a^{-3(1+\alpha)}$ dominates the energy density, leading to an effective dust-like behaviour $\rho(a) \propto a^{-3}$ in the early-time limit. In this regime, the unified matter–DE component behaves as pressureless dust, effectively mimicking cold DM and allowing for the formation of standard cosmic structures. At intermediate epochs, the expansion history encoded in $E(a)$ is governed by the competition between the matter-like term $\Omega_{m0}a^{-3(1+\alpha)}$, dominant at early times, and the constant contribution $(1-\Omega_{m0})\equiv\Omega_{DE0}$, which gradually becomes important and drives the departure from matter domination, signaling the onset of late-time accelerated phase. 
In the asymptotic future ($a\gg1$), the contribution of matter becomes negligible and Eq.~(\ref{eq:E}) approaches a constant value, implying vacuum domination and an asymptotically de Sitter expansion. A three-dimensional representation of the Hubble parameter $H(a)$ in the spatially flat $\Lambda(t)$CDM model as a function of $\alpha$ and the scale factor, $a$, is illustrated in Fig.~\ref{fig:f1}.

\section{Derivation of the total effective EoS parameter, the deceleration parameter, and the reconstruction of the DE EoS parameter}\label{section:3}

 We derived the explicit form of the total effective EoS parameter of the unified cosmic fluid in the spatially flat $\Lambda(t)$CDM model,
$w_{tot} = p_{tot}/\rho_{tot}$ (see Appendix~A for details), which characterises the combined behaviour of matter and DE in the $\Lambda(t)$CDM framework. 
The total effective EoS parameter of the unified cosmic fluid is given by (see Eq.~(\ref{eq:w_tot}))

\begin{equation*}
w_{tot}(a)=
-1+\frac{\Omega_{m0}a^{-3(1+\alpha)}}
{1-\Omega_{m0}+\Omega_{m0}a^{-3(1+\alpha)}}.
\end{equation*}

The asymptotic behaviour of $w_{tot}$ can be summarised as follows: in the early Universe, $w_{tot}(a \to 0) \to 0$; at the present epoch, $w_{tot}(a = 1) = -(1-\Omega_{m0}) = -\Omega_{DE0}$; and in the asymptotic future, $w_{tot}(a \gg 1) \to -1$. 
This sequence reproduces the standard cosmological evolution from a matter dominated era ($w_{tot} \simeq 0$) to a vacuum dominated phase, ultimately approaching an asymptotic de Sitter state ($w_{tot} \to -1$). Consequently, despite the presence of vacuum dynamics, the $\Lambda(t)$CDM model naturally reproduces the conventional dust $\rightarrow$ vacuum domination $\rightarrow$ de Sitter evolution, preserving the essential features of early time structure formation while modifying the timing of late time cosmic acceleration.

We also derived the deceleration parameter (see Appendix~A for details) and the DE EoS parameter (see Appendix~B for details), $w_{DE}$, treating DE as a distinct component in the spatially flat $\Lambda(t)$CDM model. This distinction is crucial, since $w_{tot}$ and $w_{DE}$ coincide only in the asymptotic future limit of complete DE domination, whereas at earlier times, including the present epoch, they generally differ due to the non-negligible contribution of matter. 
In the spatially flat $\Lambda(t)$CDM model, the deceleration parameter and the DE EoS parameter take, respectively, the following forms (see Eqs.~(\ref{eq:qa}) and (\ref{eq:w_DE}))

\begin{equation*}
q(a) = -1 + \frac{3}{2} \frac{\Omega_{m0} a^{-3(1+\alpha)}}{X(a)},
\end{equation*} 
\begin{equation*}
w_{DE}(a) = -\frac{1-\Omega_{ m0}a^{-3(1+\alpha)}X^{-1}(a)}{1-\Omega_{m0}a^{-3}X^{-\frac{1}{1+\alpha}}(a)},~{\rm with}~ \alpha\neq-1,
\end{equation*}
where $X(a)=1-\Omega_{m0}+\Omega_{m0}a^{-3(1+\alpha)}$. 

For $\alpha = 0$, the expression for the DE EoS parameter in Eq.~(\ref{eq:w_DE}) reduces to that of the standard $\Lambda$CDM model, with $w_{DE}(a) = -1$. At the present epoch, i.e., for $a = 1$, Eqs.~(\ref{eq:qa}) and (\ref{eq:w_DE}) yield $q(a = 1) = \tfrac{3}{2}\Omega_{m0} - 1$ and $w_{DE}(a = 1) = -1$, respectively. By contrast, the total effective EoS parameter of the unified cosmic fluid is today, according to Eq.~(\ref{eq:w_tot}), $w_{tot}(a = 1) = -\Omega_{\rm DE0}$ due to the residual matter contribution. Therefore, the $\Lambda(t)$CDM model approaches the standard $\Lambda$CDM scenario only asymptotically in the future, rather than exactly at the present epoch.

\section{Observational constraints of the $\Lambda(t)$CDM and $\Lambda$CDM models}
\label{section:4}

We performed observational constraints on the parameter sets ${\bf p}=\{H_0, \alpha, \Omega_{m0}\}$ and ${\bf p}=\{ H_0, \alpha, \Omega_{m0}, \Omega_{k0}\}$ for the spatially flat and non-flat $\Lambda(t)$CDM models, respectively, and ${\bf p}=\{H_0, \Omega_{m0}\}$ and ${\bf p}=\{H_0, \Omega_{m0}, \Omega_{k0}\}$ for the spatially flat and non-flat $\Lambda$CDM models, respectively. For this purpose, we used a compilation of 32 $H(z)$ measurements in the redshift range $z \in [0.07, 1.965]$, as listed in Table of~\citet{Cao:2023eja}, which lists data from~\citet{Simon:2004tf, Stern:2009ep, 2014RAA....14.1221Z, Ratsimbazafy:2017vga}, and~\citet{Borghi:2021rft}. Among these, 15 measurements are correlated~\citep{2012JCAP...08..006M, Moresco:2015cya, Moresco:2016mzx}. The covariance matrix for the correlated measurements is available\footnote{\url{https://gitlab.com/mmoresco/CCcovariance/}}.
We also used 13 BAO measurements from DESI DR2, 12 of which are correlated, as reported in Table IV of \citep{DESI:2025zgx}, spanning the redshift range $z \in [0.295, 2.330]$. The priors for the parameters of the $\Lambda$CDM and $\Lambda(t)$CDM models adopted in our analysis are presented in Table~\ref{table:1}.

\begin{table}[!hbp]
\caption{Priors on the model parameters.}
 \label{table:1}
 \centering
\begin{tabular}{cc}
\hline\noalign{\smallskip}
$\rm Parameter$   &     $\rm Priors$ \\
\hline
$H_0$ & $[\rm None,~ None]$ \\
$\alpha$ &  $[-0.99,1]$ \\
$\Omega_{m0}$ &  $[0.2,0.6]$  \\
$\Omega_{k0}$ &  $[-0.6,0.6]$ \\
\noalign{\smallskip}\hline
\end{tabular}
  \end{table}
 We obtained the posterior distributions of the parameters of the $\Lambda$CDM and $\Lambda(t)$CDM models using the MCMC method implemented in the Cobaya package~\citep{Torrado:2020dgo}. To produce plots and analyse the MCMC results, we applied the PYTHON GETDIST package \citep{Lewis:2019xzd}.

For model selection based on the $H(z)$ data alone, we employed the $\mathrm{AICc}$, since the number of data points is relatively small compared to the number of free parameters. The $\mathrm{AICc}$ is defined as~\citep{10.1093/biomet/76.2.297}
\begin{equation}
\mathrm{AICc} = \chi^2_{\min} + 2k + \frac{2k(k+1)}{n - k - 1},
\end{equation}
where $\chi^2_{\min}$ is the minimum chi-squared value, $k$ denotes the number of free parameters, and $n$ is the total number of data points. For the $H(z)$ data set ($n=32$, corresponding to $n/k = 6.4$–$16 < 40$), the use of $\mathrm{AICc}$ is justified, as it corrects for the known tendency of the standard $\mathrm{AIC}$ to favour overly complex models in small-sample regimes~\citep{10.1093/biomet/76.2.297}.

For the combined $H(z)$+BAO data set, the $\mathrm{AIC}$~\citep{Akaike1974} is sufficient, as the correction becomes negligible. As an additional consistency check, we compute $\mathrm{BIC}$~\citep{Schwarz1978}. The $\mathrm{AIC}$ and $\mathrm{BIC}$ can be evaluated, respectively, as
\begin{equation}
\mathrm{AIC} = \chi^2_{\min} + 2k,~~~\mathrm{BIC} = \chi^2_{\min} + k \ln n.
\end{equation}

To provide a fully Bayesian model comparison, we computed the $\mathrm{DIC}$ directly from the MCMC posterior samples~\citep{Spiegelhalter:2002,Gelman:2014}. Unlike approximate expressions based solely on best-fit values, this approach accounts for the full shape of the posterior distributions and parameter degeneracies. The $\mathrm{DIC}$ is defined as
\begin{equation}
\mathrm{DIC} = D(\bar{\theta}) + 2p_D,
\end{equation}
where $D(\theta) = -2\ln \mathcal{L}(\theta)$ is the deviance, $\mathcal{L}(\theta)$ is the likelihood function of the data given the model parameters, and $\bar{\theta}$ denotes the posterior mean of the parameters. The effective number of parameters is given by
\begin{equation}
p_D = \langle D(\theta) \rangle - D(\bar{\theta}),
\end{equation}
where the angle brackets denote the average over the MCMC posterior samples.

Furthermore, we evaluated the Akaike weights $w_i$~\citep{burnham2002}
\begin{equation}
w_i = \frac{\exp(-\Delta_i/2)}{\sum_{j=1}^{R} \exp(-\Delta_j/2)},
\end{equation}
where $\Delta_i = \mathrm{AICc}_i - \min(\mathrm{AICc})$ (or $\Delta_i = \mathrm{AIC}_i - \min(\mathrm{AIC})$), and $R$ denotes the total number of models under consideration. These weights can be interpreted as the relative probability that a given model provides the best description of the data, in the sense of minimising information loss, assuming that one of the candidate models is correct.

In addition, the quantities $\Delta\chi^2_{\min} = \chi^2_{\min,i} - \min(\chi^2_{\min})$ and the reduced chi-squared, $\chi^2_{\min}/{\mathrm {dof}}$, are used to assess the goodness of fit of each model, where ${\mathrm {dof}} = n - k$ is the number of degrees of freedom.
The comparison of the models is performed using the differences $\Delta \mathrm{AICc}$ (or $\Delta \mathrm{AIC}$), $\Delta \mathrm{BIC}$, and $\Delta \mathrm{DIC}$, defined relative to the model that minimises the corresponding information criterion. All values of $\Delta \chi^2_{\min}$, $\Delta \mathrm{AICc}$, $\Delta \mathrm{AIC}$, $\Delta \mathrm{BIC}$, and $\Delta \mathrm{DIC}$ are computed with respect to the best-performing model within each data set separately. Akaike weights are then used to quantify the relative statistical support of the competing models.

\section{Results and discussion}\label{section:5}
\subsection{Properties of the spatially flat $\Lambda(t)$CDM model}

\subsubsection{Expansion of the Universe}
\begin{figure}[ht!]
\begin{center}
\includegraphics[width=\columnwidth]{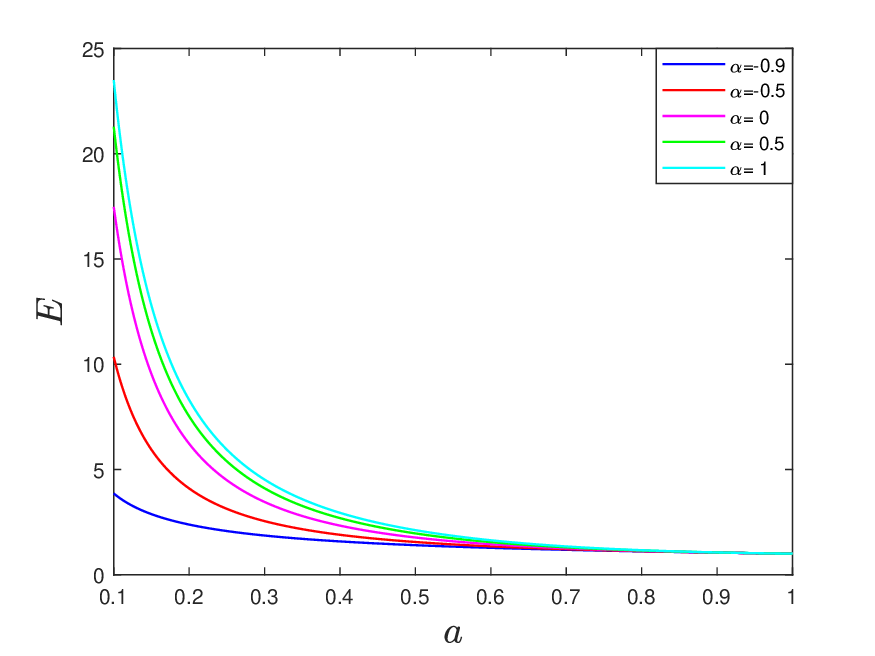}
\end{center}
 \caption {Normalised Hubble parameter in the spatially flat $\Lambda(t)$CDM model as a function of the scale factor for different values of the vacuum dynamics parameter, $\alpha$.}
 \label{fig:f2}
 \end{figure}

We numerically solved the first Friedmann equation, Eq.~(\ref{eq:E}), without spatial curvature and radiation terms, to study the expansion of the Universe in the spatially flat, radiation-free $\Lambda(t)$CDM model. The evolution of the normalised Hubble parameter as a function of the scale factor for different values of the vacuum dynamics parameter, $\alpha$, is shown in Fig.~\ref{fig:f2}. This quantity depends sensitively on $\alpha$. For negative values of $\alpha$, corresponding to effective matter creation, the matter component dilutes more slowly than in the standard $\Lambda$CDM model, leading to a suppressed expansion rate at early times. For negative values of $\alpha$, corresponding to effective matter creation, the matter component dilutes more slowly than in the standard $\Lambda$CDM model, leading to a suppressed expansion rate at early times. In contrast, for positive values of $\alpha$, associated with effective matter annihilation, the matter density decreases faster, resulting in an enhanced expansion rate relative to the $\Lambda$CDM model. All trajectories converge at late times due to the normalisation condition $E(a = 1) = 1$.

\subsubsection{Analysis of the total effective EoS parameter and the deceleration parameter}
\begin{figure}[ht!]
\begin{center}
\includegraphics[width=\columnwidth]{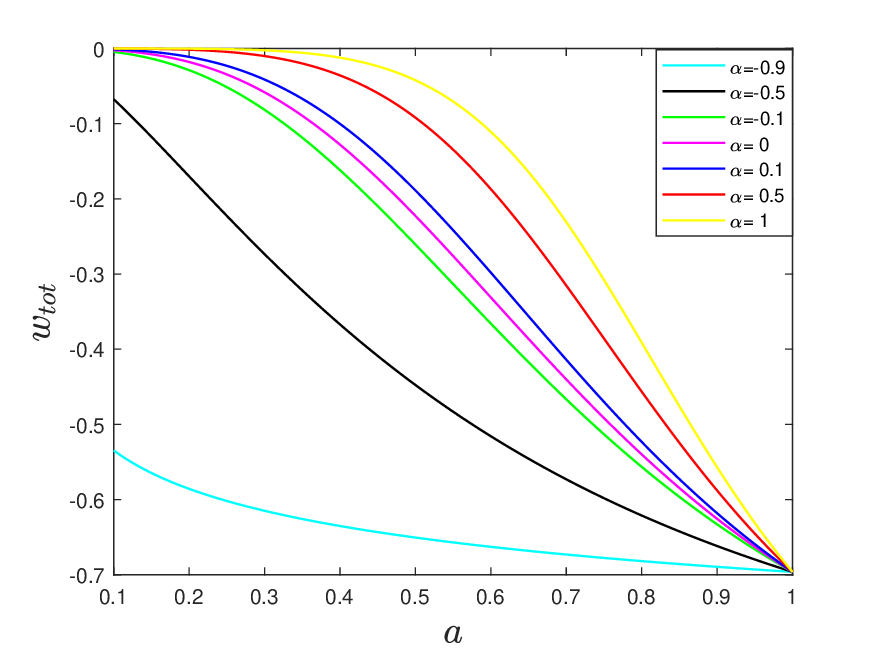}
\end{center}
 \caption {Total effective EoS parameter of the unified cosmic fluid as a function of the scale factor for different values of the vacuum dynamics parameter, $\alpha$, in the spatially flat $\Lambda(t)$CDM model.}
 \label{fig:f3}
 \end{figure}

The numerical solution of Eq.~(\ref{eq:w_tot}) for the total effective EoS parameter of the unified cosmic fluid in the spatially flat $\Lambda(t)$CDM model for different values of the vacuum dynamics parameter, $\alpha$, in the scale factor range $a \in [0.1,1]$ is shown in Fig.~\ref{fig:f3}.

The evolution of the total effective EoS parameter $w_{tot}(a)$ shows that, at early times ($a \ll 1$), the unified cosmic fluid behaves as pressureless dust, with $w_{tot}(a) \simeq 0$, independently of the value of $\alpha$. As the Universe expands, $w_{tot}(a)$ gradually decreases and enters the quintessence regime, $-1 < w_{tot}(a) < -1/3$~\citep{Caldwell:2005tm, Peebles:2002gy}. It does not cross the phantom divide $w = -1$~\citep{Caldwell1999, Elizalde2004}, even for negative values of $\alpha$, indicating that the total cosmic fluid remains non-phantom throughout the evolution. We also find that increasing $\alpha$ delay the transition from the matter-dominated epoch to the DE-dominated epoch. In particular, larger $\alpha$ shifts both the departure from matter-like behaviour ($w_{tot} \simeq 0$) and the onset of negative pressure to later times. 
For $\alpha = 0$, the total effective EoS parameter evolves as
\begin{equation*}
w_{tot}(a)=
-1+\frac{\Omega_{m0}a^{-3}}
{1-\Omega_{m0}+\Omega_{m0}a^{-3}}.
\end{equation*}

All $w_{tot}$ trajectories, independently of the value of $\alpha$, converge at the present epoch, yielding $w_{tot}(a = 1) = \Omega_{m0} - 1.$ In our analysis, assuming $\Omega_{m0} = 0.3038$~\citep{DESI:2025hao}, this yields $w_{tot}(a = 1) = -0.6962$.

\begin{figure}[ht!]
\begin{center}
\includegraphics[width=\columnwidth]{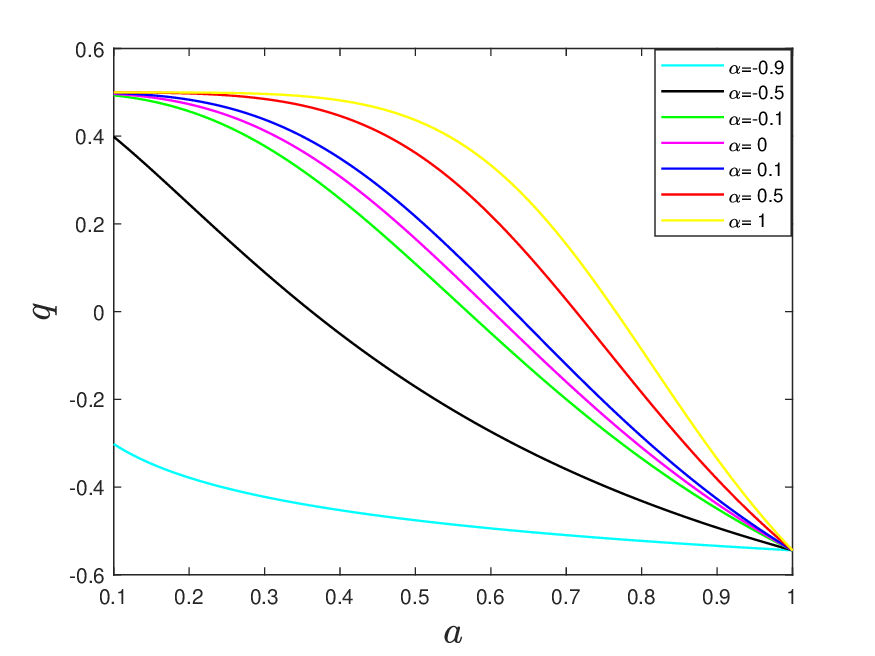}
\end{center}
 \caption {Evolution of the deceleration parameter as a function of the scale factor for different values of the vacuum dynamics parameter, $\alpha$, in the spatially flat $\Lambda(t)$CDM model.}
 \label{fig:f4}
 \end{figure}

The numerical solution of Eq.~(\ref{eq:qa}) for the reconstructed deceleration parameter, $q(a)$, in the spatially flat $\Lambda(t)$CDM model for various values of the vacuum dynamics parameter, $\alpha$, over the scale factor range $a \in [0.1,1]$ is presented in Fig.~\ref{fig:f4}. At early times, $q(a) > 0$, indicating a decelerated expansion dominated by matter. As the vacuum component becomes dynamically significant, $q(a)$ decreases and eventually crosses zero, marking the transition from decelerated to accelerated expansion of the Universe. Our results indicate that increasing $\alpha$ shifts this transition to later epochs, implying that a stronger vacuum–matter interaction delays the onset of cosmic acceleration. Despite these differences in evolutionary trajectories, all solutions converge to the same current value of the deceleration parameter, $q_0 = -0.5443$, corresponding to $\Omega_{m0}=0.3038$~\citep{DESI:2025hao}.

\subsection{Observational constraints of the $\Lambda(t)$CDM and $\Lambda$CDM models}

\begin{figure*}[!htbp]
\centering
\begin{minipage}{.49\textwidth}
\includegraphics[width=\linewidth]{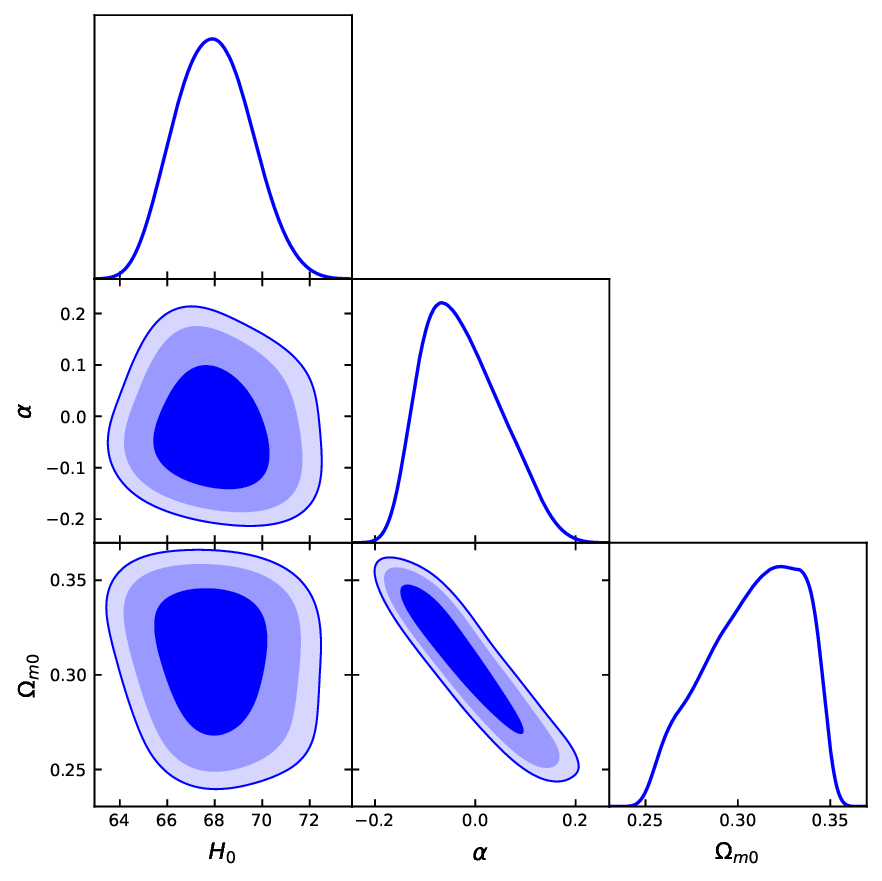}
\end{minipage}
\hfill
\begin{minipage}{.49\textwidth}
\includegraphics[width=\linewidth]{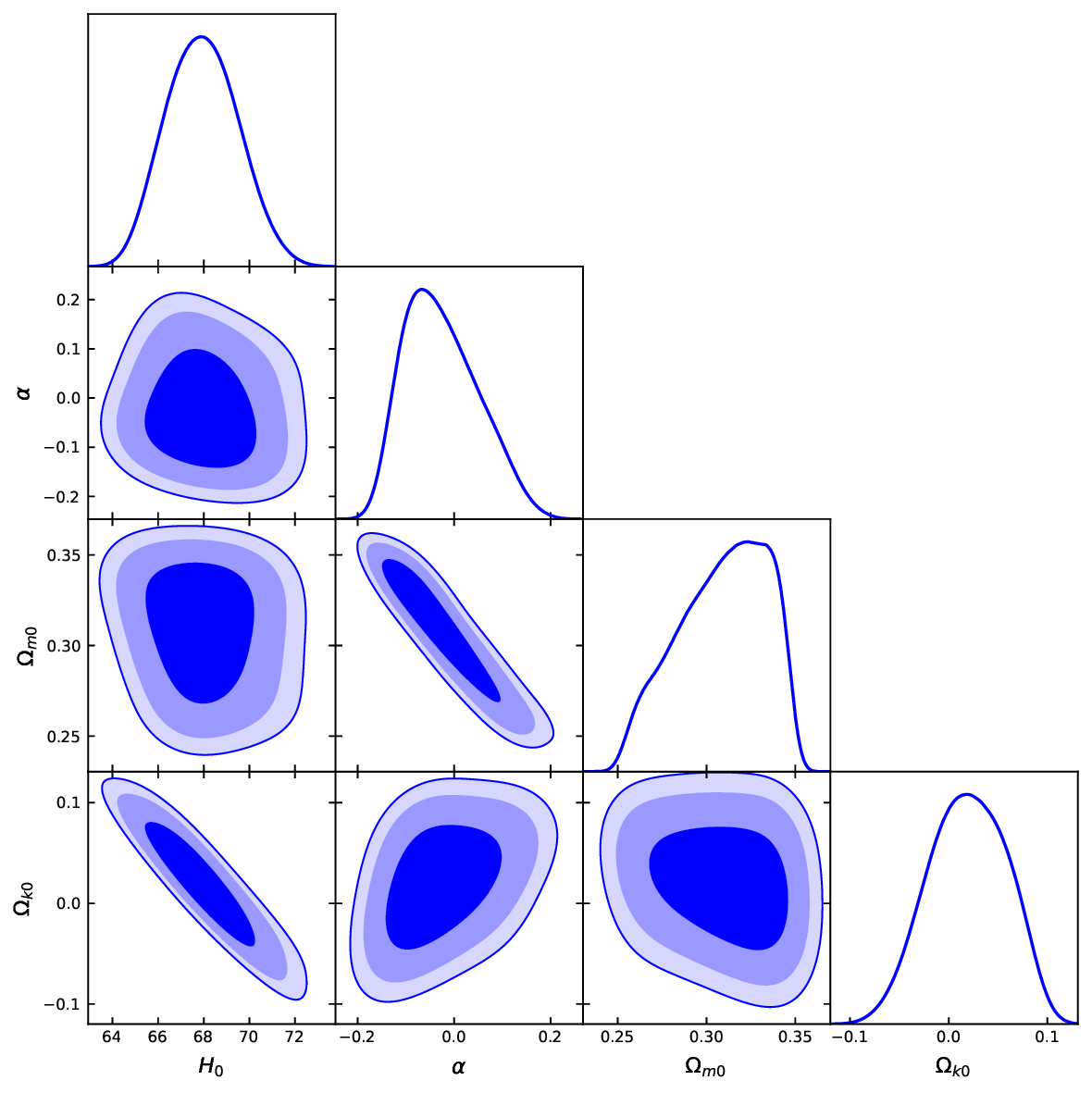}
\end{minipage}
\caption{One-dimensional likelihoods and 1$\sigma$, 2$\sigma$, and 3$\sigma$ confidence level contours for the parameter constraints of the $\Lambda(t)$CDM model from $H(z)$ measurements: for the flat space (left panel) and the non-flat space (right panel).}
\label{fig:f5}
\end{figure*}

\begin{figure*}[!htbp]
\centering
\begin{minipage}{.49\textwidth}
\includegraphics[width=\linewidth]{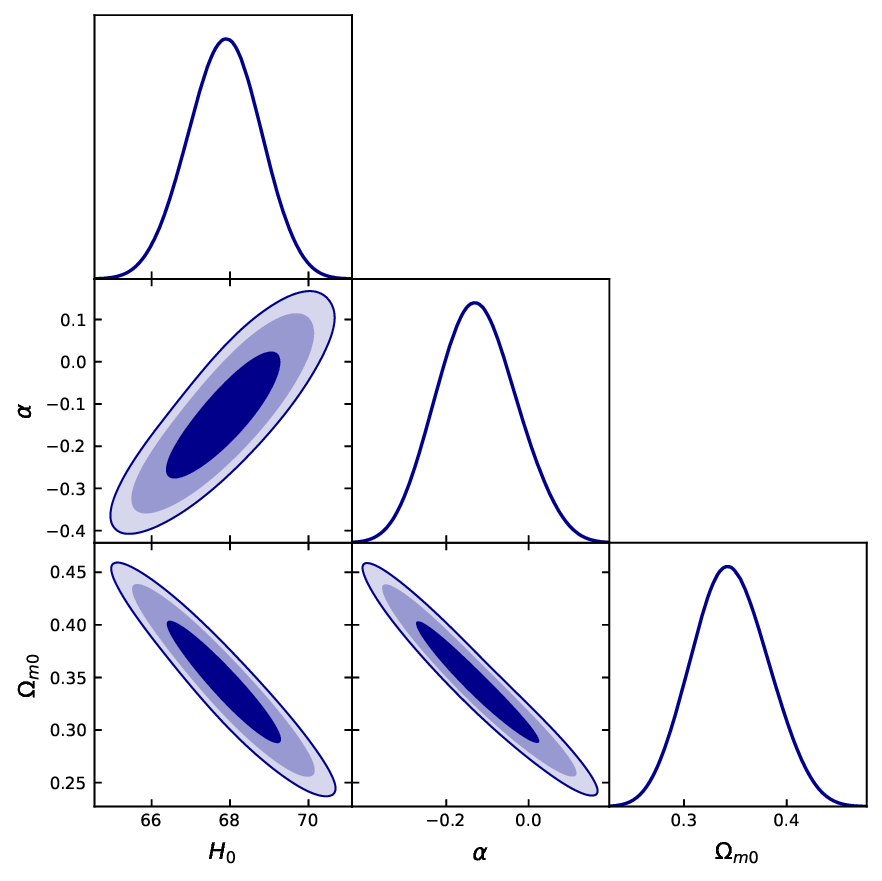}
\end{minipage}
\hfill
\begin{minipage}{.49\textwidth}
\includegraphics[width=\linewidth]{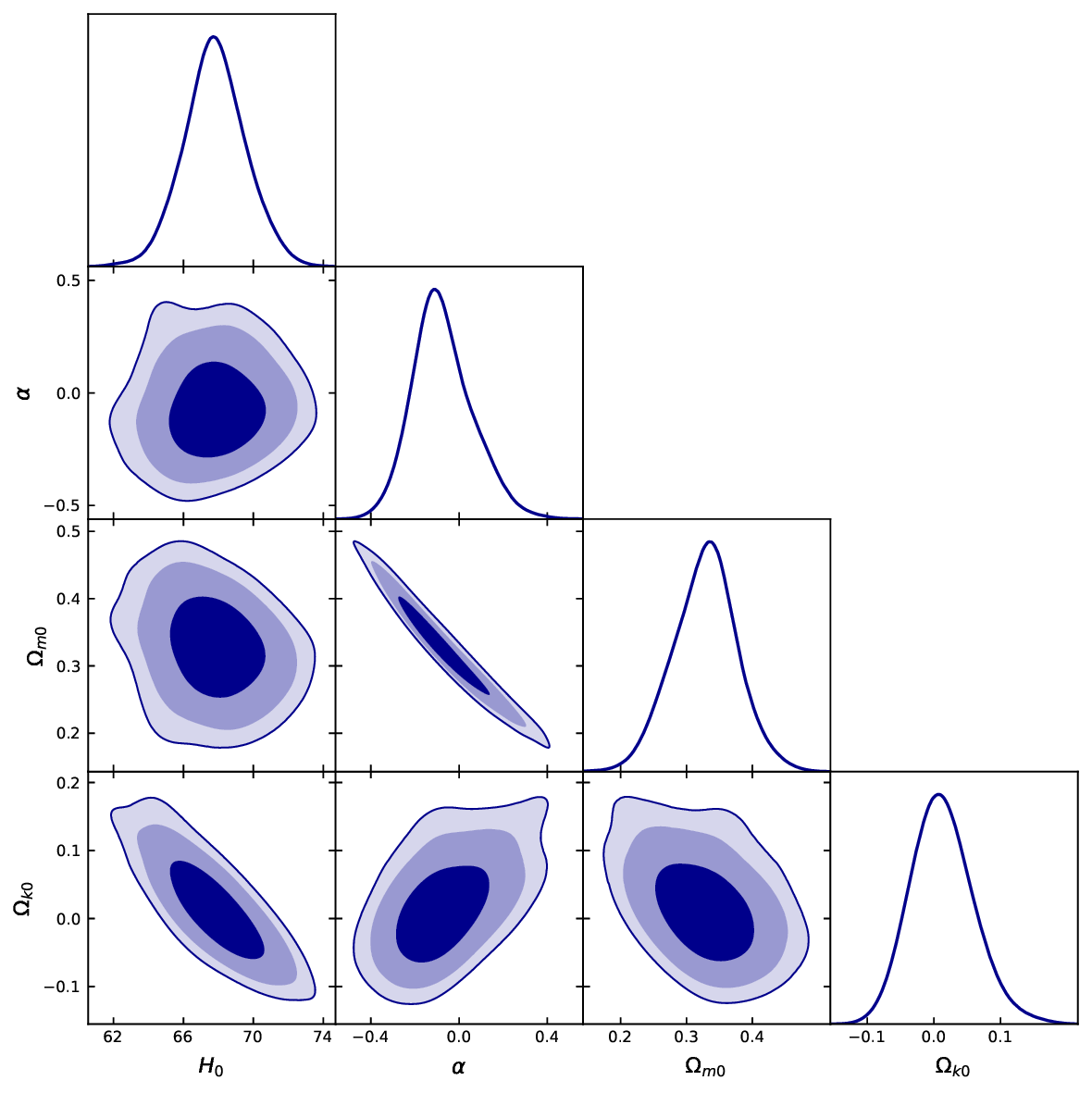}
\end{minipage}
\caption{One-dimensional likelihoods and 1$\sigma$, 2$\sigma$, and 3$\sigma$ confidence level contours for the parameter constraints of the $\Lambda(t)$CDM model from the combination of $H(z)$ + BAO measurements: for the flat space (left panel) and the non-flat space (right panel).}
\label{fig:f6}
\end{figure*}

\begin{figure*}[!htbp]
\centering
\begin{minipage}{.49\textwidth}
\includegraphics[width=\linewidth]{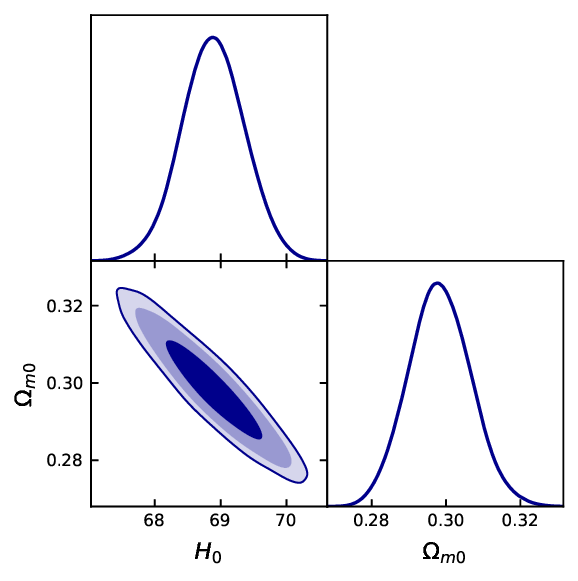}
\end{minipage}
\hfill
\begin{minipage}{.49\textwidth}
\includegraphics[width=\linewidth]{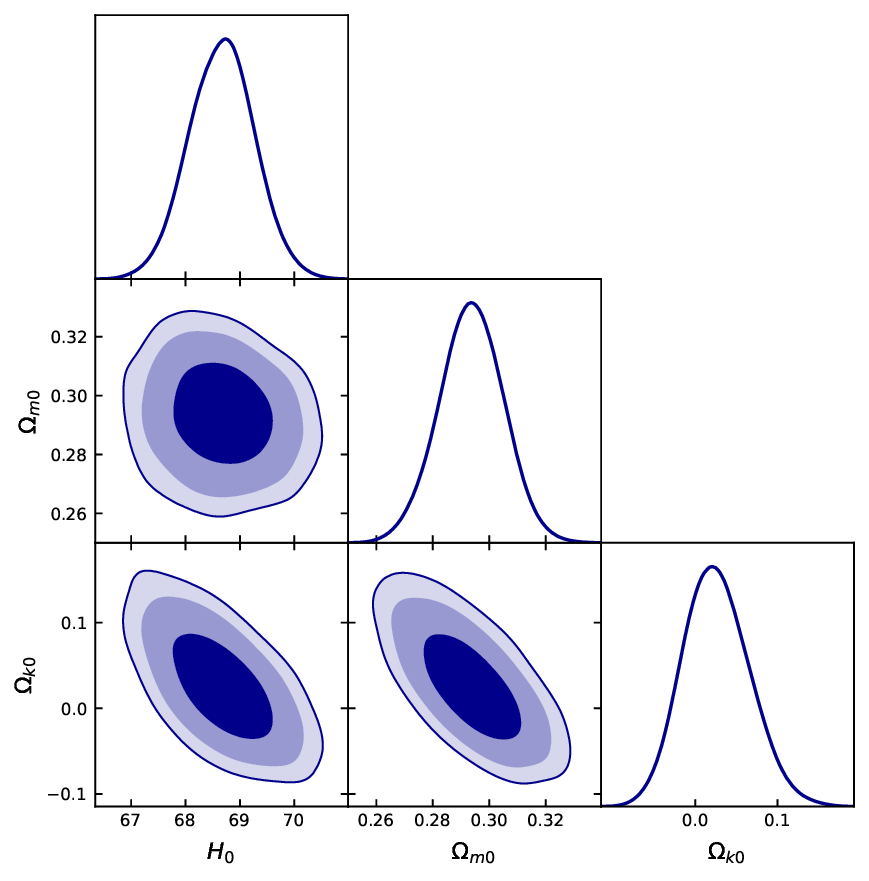}
\end{minipage}
\caption{One-dimensional likelihoods and 1$\sigma$, 2$\sigma$, and 3$\sigma$  confidence level contours for the parameter constraints on the parameters of the $\Lambda$CDM model from the combination of $H(z)$ + BAO measurements: for the flat space (left panel) and the non-flat space (right panel).}
\label{fig:f7}
\end{figure*}

Based on the MCMC analysis, we obtained the one-dimensional likelihood distributions and the 1$\sigma$, 2$\sigma$, and 3$\sigma$ confidence contours for the model parameters using 32 $H(z)$ measurements for the spatially flat $\Lambda(t)$CDM model (left panel of Fig.~\ref{fig:f5}) and the spatially non-flat $\Lambda(t)$CDM model (right panel of Fig.~\ref{fig:f5}). Using the combined data set of 32 $H(z)$ and 13 BAO measurements, we further derived the joint 1$\sigma$, 2$\sigma$, and 3$\sigma$ confidence contours for the spatially flat and non-flat $\Lambda(t)$CDM models (Fig.~\ref{fig:f6}), as well as for the standard spatially flat $\Lambda$CDM model and its spatially non-flat extension (Fig.~\ref{fig:f7}).

The one-dimensional posterior median values and their corresponding $68\%$ credible intervals for the spatially flat and non-flat $\Lambda$CDM and $\Lambda(t)$CDM models, obtained from various combinations of $H(z)$ and $H(z)$+BAO data, are summarised in Table~\ref{table:2}. We report median values instead of means, as they provide a more robust estimate in the presence of non-Gaussian and asymmetric posterior distributions. Using $H(z)$ data alone results in relatively large uncertainties in all models, particularly for the spatially non-flat $\Lambda(t)$CDM scenario.
For this model, we find $H_0 = 61.08^{+9.60}_{-6.38}~{\rm km~s^{-1}Mpc^{-1}}$, $\Omega_{m0} = 0.214^{+0.237}_{-0.173}$, $\Omega_{k0} = 0.236^{+0.253}_{-0.274}$, and $\alpha = 0.475^{+1.633}_{-0.835}$, indicating weak constraints and strong parameter degeneracies. In contrast, the spatially flat $\Lambda(t)$CDM model and both spatially flat and non-flat $\Lambda$CDM models yield $H_0 \sim 67$–$68~{\rm km~s^{-1}Mpc^{-1}}$, although with broader uncertainties than those obtained from the combined data sets, while the parameter $\alpha$ remains statistically consistent with zero in both spatially flat and non-flat $\Lambda(t)$CDM models.
The inclusion of BAO data significantly tightens the parameter constraints for all models (see Table~\ref{table:2}). For the spatially non-flat $\Lambda(t)$CDM model, we find $H_0 = 67.76 \pm 1.66~{\rm km~s^{-1}Mpc^{-1}}$, $\Omega_{m0} = 0.328^{+0.049}_{-0.045}$, $\Omega_{k0} = 0.009^{+0.050}_{-0.044}$, and $\alpha = -0.081 \pm 0.133$, favouring a spatially flat Universe with a matter density parameter, $\Omega_{m0}$, compatible with that of the standard $\Lambda$CDM model.
Similarly, the spatially flat $\Lambda(t)$CDM model yields $H_0 = 67.87 \pm 0.91~{\rm km~s^{-1}Mpc^{-1}}$, $\Omega_{m0} = 0.344^{+0.040}_{-0.035}$, and $\alpha = -0.127 \pm 0.097$, with BAO data substantially reducing the uncertainties in $\alpha$ while maintaining consistency with zero. In both spatially flat and non-flat $\Lambda$CDM models, BAO data reduce the uncertainties in $H_0$ and $\Omega_{m0}$ by a factor of $3 - 6$ and constrain the curvature parameter, $\Omega_{k0}$, close to zero; in the spatially non-flat $\Lambda$CDM scenario, $\Omega_{k0} = 0.024^{+0.043}_{-0.038}$, again favouring a nearly flat geometry.
Consequently, the inclusion of BAO data reduces the Hubble tension with the $Planck$ CMB result\footnote{We emphasise that the quoted $Planck$ CMB constraint on $H_0$ is obtained under the assumption of the standard $\Lambda$CDM cosmological model. A dedicated CMB analysis within the $\Lambda(t)$CDM framework would, in general, yield different parameter estimates and could alter the level of statistical agreement with the $Planck$ CMB result. Such an analysis is beyond the scope of the present work, and the $Planck$ CMB constraint is therefore used here solely as a reference benchmark.} to below the $1\sigma$ level for all models, while the tension with SH0ES is reduced to $\sim 2.6$–$3.2\sigma$.
Thus, the inclusion of BAO data substantially reduces parameter degeneracies and leads to mutually consistent cosmological parameter estimates across all considered models, while the vacuum dynamics parameter $\alpha$ remains statistically consistent with zero in both spatially flat and non-flat $\Lambda(t)$CDM models.

We evaluated the performance of the spatially flat and non-flat $\Lambda(t)$CDM and $\Lambda$CDM models using goodness-of-fit statistics and information criteria based on the $H(z)$ and combined $H(z)$+BAO data sets (see Tables~\ref{table:3}--\ref{table:4}). The results for the $H(z)$ data are summarised in Table~\ref{table:3}.
All models provide an excellent fit to the data, with $\chi^2_{\min}/\mathrm{dof} \approx 0.5$, indicating no significant tension between the models and the observations. When using $H(z)$ data alone, the $\Lambda(t)$CDM models yield less tightly constrained cosmological parameters compared to the standard $\Lambda$CDM model, particularly in the spatially non-flat case. The vacuum dynamics parameter, $\alpha$, is consistent with zero within $1\sigma$, suggesting no statistically significant deviation from the standard $\Lambda$CDM scenario.
The spatially non-flat $\Lambda(t)$CDM model yields the minimum $\chi^2$; however, the improvement is marginal ($\Delta\chi^2_{\min} \lesssim 1$), and all models perform at a statistically comparable level.
When model complexity is taken into account, the information criteria ($\mathrm{AICc}$ and $\mathrm{BIC}$) favour the spatially flat $\Lambda$CDM model due to its smaller number of free parameters. Nevertheless, the differences with respect to the extended models remain small, indicating no strong statistical preference.
A similar conclusion is reached from the $\mathrm{DIC}$ analysis, where all models show differences within $\Delta \mathrm{DIC} \lesssim 2$, implying statistical equivalence. In particular, the spatially flat $\Lambda(t)$CDM model does not exhibit a significant advantage over the standard $\Lambda$CDM scenario.
The effective number of parameters $p_D$ suggests that the additional degrees of freedom in the $\Lambda(t)$CDM models are not fully constrained by the $H(z)$ data, pointing to mild parameter degeneracies rather than an improvement in model performance.

The inclusion of BAO data (Table~\ref{table:4}) leads to tighter parameter constraints, as reflected in higher values of $p_D$. However, the overall model ranking remains unchanged. The spatially flat $\Lambda$CDM model is mildly preferred by both $\mathrm{AIC}$ and $\mathrm{BIC}$, while the extended models remain statistically competitive, with small differences in both $\Delta \mathrm{AIC}$ and $\Delta \mathrm{DIC}$.

Overall, the combined $H(z)$ and BAO data do not provide strong statistical evidence in favour of dynamical vacuum models. The standard spatially flat $\Lambda$CDM cosmology remains a robust and fully consistent description of the data within the considered framework.

\begin{table*}[!ht]
\centering
\begin{threeparttable}
\caption{Posterior constraints on the model parameters.}
\label{table:2}
\begin{tabular}{cccccc}
\hline\noalign{\smallskip}
Model & Data & {$H_0$\tnote{a}} & $\Omega_{m0}$ & $\Omega_{k0}$ & $\alpha$ \\
\hline \noalign{\smallskip}
Flat $\Lambda(t)$CDM & $H(z)$ & $67.91\pm1.62$ & $0.312^{+0.024}_{-0.031}$ & $-$ & $-0.035^{+0.093}_{-0.071}$ \\
\noalign{\smallskip}
 & $H(z)$+BAO & $67.87\pm0.91$ & $0.344^{+0.040}_{-0.035}$ & $-$ & $-0.127\pm0.097$ \\
\hline \noalign{\smallskip}
Non-flat $\Lambda(t)$CDM & $H(z)$ & $61.08^{+9.60}_{-6.38}$ & $0.214^{+0.237}_{-0.173}$ & $0.236^{+0.253}_{-0.274}$ & $0.475^{+1.633}_{-0.835}$ \\
\noalign{\smallskip}
 & $H(z)$+BAO & $67.76\pm1.66$ & $0.328^{+0.049}_{-0.045}$ & $0.009^{+0.050}_{-0.044}$ & $-0.081\pm0.133$ \\
 \hline \noalign{\smallskip}
Flat $\Lambda$CDM & $H(z)$ & $67.66\pm 3.05$ & $0.326^{+0.067}_{-0.054}$ & $-$ & $-$ 
\\ \noalign{\smallskip}
 & $H(z)$+BAO & $68.90\pm0.49$ & $0.298\pm0.009$ & $-$ & $-$ \\
\hline \noalign{\smallskip}
Non-flat $\Lambda$CDM & $H(z)$ & $67.99\pm3.83$ & $0.350^{+0.124}_{-0.133}$ & $-0.062^{+0.404}_{-0.354}$ & $-$ \\
\noalign{\smallskip}
 & $H(z)$+BAO & $68.68\pm0.60$ & $0.294\pm0.011$ & $0.024^{+0.043}_{-0.038}$ & $-$ \\
\noalign{\smallskip}\hline
\end{tabular}
\tablefoot{
(a) km\,s$^{-1}$\,Mpc$^{-1}$;
(b) We report posterior median values and 68\% credible intervals for spatially flat and non-flat $\Lambda(t)$CDM and $\Lambda$CDM models obtained from the $H(z)$ and $H(z)$+BAO data sets.
}
\end{threeparttable}
\end{table*}

\begin{table*}[!ht]
\centering
\begin{threeparttable}
\caption{Model selection statistics for the $H(z)$ data.}
\label{table:3}
\begin{tabular}{cccccccccccccc}
\hline\noalign{\smallskip}
$\rm Model$ & $ \rm \chi^2_{\min}$ & $\Delta\chi^2_{\min}$ & ${\mathrm {dof}}$ & $\rm \chi^2_{\min}/{\mathrm {dof}}$ & $\mathrm{AICc}$ & $\mathrm{BIC}$ & $\mathrm{DIC}$& $p_D$ & $\Delta \mathrm{AICc}$ & $\Delta \mathrm{BIC}$ & $\Delta \mathrm{DIC}$ & $w_i$ \\
\hline \noalign{\smallskip}
Flat $\Lambda(t)$CDM & $14.68$ & $0.57$ & $29$ & $0.506$ & $21.54$ & $25.08$ & $17.42$ & $1.37$ & $2.39$ & $3.42$ & $0.00$ & $0.177$\\

Non-flat $\Lambda(t)$CDM & $14.11$ & $0.00$ & $28$ & $0.504$ & $23.59$ & $27.97$ & $19.95$ & $2.91$ & $4.45$ & $6.31$ & $2.53$ & $0.063$\\
  
Flat $\Lambda$CDM & $14.73$ & $0.62$ & $30$ & $0.491$ & $19.14$ & $21.66$ & $18.49$ & $1.88$ & $0.00$ & $0.00$ & $1.07$ & $0.584$\\

Non-flat $\Lambda$CDM & $14.69$ & $0.58$ & $29$ & $0.507$ & $21.55$ & $25.09$ & $19.20$ & $2.26$ & $2.40 $ & $3.43$ & $1.78$ & $0.176$ \\
\noalign{\smallskip}\hline
\end{tabular}
\tablefoot{
$\Delta \mathrm{AICc}$, $\Delta \mathrm{BIC}$, and $\Delta \mathrm{DIC}$ are computed with respect to the model with the minimum value of each criterion. The quantity $p_D$ denotes the effective number of parameters entering the $\mathrm{DIC}$.
}
\end{threeparttable}
\end{table*}

\begin{table*}[!ht]
\centering
\begin{threeparttable}
\caption{Model selection statistics for the $H(z)$+BAO data.}
\label{table:4}
\begin{tabular}{ccccccccccccc}
\hline\noalign{\smallskip}
$\rm Model$ & $ \rm \chi^2_{\min}$ & $\Delta\chi^2_{\min}$ & ${\mathrm {dof}}$ & $\rm \chi^2_{\min}/{\mathrm {dof}}$ & $\mathrm{AIC}$ & $\mathrm{BIC}$ & $\mathrm{DIC}$& $p_D$ & $\Delta \mathrm{AIC}$ & $\Delta \mathrm{BIC}$ & $\Delta \mathrm{DIC}$ & $w_i$ \\
\hline \noalign{\smallskip}
Flat $\Lambda(t)$CDM & $24.44$ & $0.00$ & $42$ & $0.582$ & $30.44$ & $35.86$ & $29.98$ & $2.77$ & $0.99 $ & $2.80$ & $0.50$ & $0.269 $ \\

Non-flat $\Lambda(t)$CDM & $24.47$ & $0.03 $ & $41$ & $0.597$ & $32.47$ & $39.70$ & $32.33$ & $3.91$ & $3.02$ & $6.63$ & $2.85$ & $0.097 $\\
  
Flat $\Lambda$CDM & $25.45$ & $1.01$ & $43$ & $0.592$ & $29.45$ & $33.06$ & $29.48$ & $2.02$ & $0.00 $ & $0.00 $ & $0.00$ & $0.441$\\

Non-flat $\Lambda$CDM & $25.10$ & $0.66$ & $42$ & $0.598$ & $31.10$ & $36.52$ & $31.14$ & $3.02$ & $1.65$ & $3.46 $ & $1.66$ & $0.193$ \\
\noalign{\smallskip}\hline
\end{tabular}
\tablefoot{
$\Delta \mathrm{AIC}$, $\Delta \mathrm{BIC}$, and $\Delta \mathrm{DIC}$ are computed with respect to the model with the minimum value of each criterion. The quantity $p_D$ denotes the effective number of parameters entering the $\mathrm{DIC}$.
}
\end{threeparttable}
\end{table*}

\section{Conclusions}\label{section:6}
We analysed the dynamical behaviour of the spatially flat $\Lambda(t)$CDM model by studying the expansion history of the Universe through the normalised Hubble  parameter, $E(a)$, including the total effective EoS parameter of the unified cosmic fluid, $w_{tot}(a)$, and the deceleration parameter, $q(a)$.  The vacuum dynamics parameter, $\alpha$, modifies the expansion history relative to the standard $\Lambda$CDM model, with matter creation ($\alpha<0$) leading to a suppressed expansion rate and matter annihilation ($\alpha>0$) resulting in an enhanced expansion rate. Despite these modifications, the total effective EoS parameter does not cross the phantom divide, ensuring a physically consistent evolution.
The total effective EoS parameter exhibits a universal behaviour: at early times, the unified cosmic fluid behaves as pressureless dust ($w_{tot} \simeq 0$), while at later times it evolves into the quintessence regime ($-1 < w_{tot} < -1/3$). The phantom divide $w = -1$ is not crossed for any value of the parameter $\alpha$, ensuring a non-phantom cosmic evolution at the level of the total fluid.
The parameter $\alpha$ controls the timing of the transition between cosmological epochs. Larger values of $\alpha$ delays the transition from the matter-dominated to the DE-dominated epoch, shifting the onset of negative values of $w_{tot}$ to later times. Despite these differences, all solutions converge at the present epoch, yielding $w_{tot}(a = 1) = \Omega_{m0} - 1$.
A similar behaviour is found for the deceleration parameter: the Universe evolves from an early decelerated phase ($q > 0$) to a late-time accelerated phase ($q < 0$). Larger values of $\alpha$ delay the transition to acceleration, while all models converge to the same present-day value, $q(a = 1) = \frac{3}{2}\Omega_{m0} - 1$.
Despite these late-time modifications, the $\Lambda(t)$CDM model preserves the standard matter-like behaviour at early times, ensuring consistency with structure formation. Overall, the $\Lambda(t)$CDM model represents a viable extension of the standard $\Lambda$CDM framework, in which vacuum dynamics shifts the epoch of cosmic acceleration without altering the early-time evolution of the Universe.

Based on the MCMC analysis, we constrained the parameters of both spatially flat and non-flat $\Lambda(t)$CDM models using 32 $H(z)$ measurements from cosmic chronometers ($z \in [0.07, 1.965]$) and 13 BAO measurements from DESI DR2 ($z \in [0.295, 2.330]$). The one-dimensional likelihoods and $1\sigma$, $2\sigma$, and $3\sigma$ confidence contours from $H(z)$ alone and in combination with the $H(z)$+BAO data are shown in Figs.~\ref{fig:f5}–\ref{fig:f7}. Using $H(z)$ data alone leads to large uncertainties, particularly in the spatially non-flat $\Lambda(t)$CDM model, where strong parameter degeneracies are present. The vacuum dynamics parameter, $\alpha$, remains statistically consistent with zero in all $\Lambda(t)$CDM scenarios.

The inclusion of BAO data substantially tightens parameter constraints for all models. For the spatially non-flat $\Lambda(t)$CDM model, we find $H_0 = 67.76 \pm 1.66~{\rm km~s^{-1}Mpc^{-1}}$, $\Omega_{m0} = 0.328^{+0.049}_{-0.045}$, $\Omega_{k0} = 0.009^{+0.050}_{-0.044}$, and $\alpha = -0.081 \pm 0.133$, favouring a nearly spatially flat Universe with a matter density parameter, $\Omega_{m0}$, compatible with the standard $\Lambda$CDM model. The spatially flat $\Lambda(t)$CDM model yields similar values with slightly tighter constraints. In both spatially flat and non-flat models, the BAO data reduce the uncertainties in $H_0$ and $\Omega_{m0}$ by a factor of $3 - 6$ and constrain the curvature parameter, $\Omega_{k0}$, close to zero. Consequently, the Hubble tension with $Planck$ CMB measurements is reduced to below the $1\sigma$ level, while the tension with SH0ES decreases to $\sim 2.6-3.2\sigma$.

We performed a comparative analysis of spatially flat and non-flat $\Lambda(t)$CDM and $\Lambda$CDM models using $H(z)$ and $H(z)$ + BAO data.
In this analysis, all considered models provide an excellent fit to the $H(z)$ and combined $H(z)$+BAO data, with $\chi^2_{\min}/\mathrm{dof} \approx 0.5$, indicating no significant tension with observations. The $\Lambda(t)$CDM models yield weaker parameter constraints, and the vacuum dynamics parameter, $\alpha$, remains consistent with zero, showing no statistically significant deviation from the standard $\Lambda$CDM scenario.
Although the spatially non-flat $\Lambda(t)$CDM model attains the lowest $\chi^2_{\min}$, the improvement is marginal ($\Delta\chi^2_{\min} \lesssim 1$), and all models perform at a statistically comparable level. Information criteria ($\mathrm{AICc}$, $\mathrm{BIC}$, and $\mathrm{DIC}$), which penalize model complexity, consistently favour the spatially flat $\Lambda$CDM model. The differences remain small ($\Delta \mathrm{AIC}$, $\Delta \mathrm{BIC}$, $\Delta \mathrm{DIC} \lesssim 2$), indicating no strong statistical preference; however, the Akaike weights assign the largest statistical support to the spatially flat $\Lambda$CDM model (about $44$–$58\%$), while the extended $\Lambda(t)$CDM models receive substantially lower support due to the additional parameter $\alpha$.
The effective number of parameters suggests that the extra degrees of freedom in the $\Lambda(t)$CDM models are not fully constrained, pointing to mild parameter degeneracies rather than improved performance. The inclusion of BAO data reduces these degeneracies and drives all models towards a spatially flat Universe with parameters consistent with the standard $\Lambda$CDM model.

Overall, current data do not provide compelling evidence in favour of dynamical vacuum models, and the standard spatially flat $\Lambda$CDM model remains a robust and consistent description of the observations.

\begin{acknowledgements}  
The author thanks Øyvind Grøn for a careful reading of the manuscript, for valuable comments and corrections, and for improvements that clarified the presentation of Appendices A and B. The author also thanks the anonymous referee for helpful and insightful comments that have significantly improved the clarity and overall quality of this work.
\end{acknowledgements}

\appendix
\section{Derivation of the deceleration parameter and total effective EoS parameter}
\label{app:A}

\subsection{Derivation of the deceleration parameter}

The deceleration parameter is defined as
\begin{equation}\label{eq:q1}
q(a)\equiv -\frac{\ddot{a}}{aH^2(a)},
\end{equation}
which can be rewritten as
\begin{equation}\label{eq:q_a}
q(a) = -1 - a \frac{d \ln H(a)}{da},
\end{equation}
where $a$ is the scale factor. The over-dot denotes the derivative with respect to cosmic time.

Using Eq.~(\ref{eq:E}), where $E(a) \equiv H(a)/H_0$ and
\begin{equation}
E(a)^2 = X^{\frac{1}{1+\alpha}}(a),
\end{equation}
where
\begin{equation}\label{eq:X}
X(a)=(1-\Omega_{m0})+\Omega_{m0}a^{-3(1+\alpha)},
\end{equation}
we can write 
\begin{equation}
\ln H(a) = \ln H_0 + \ln E(a)
= \ln H_0 + \frac{1}{2(1+\alpha)} \ln X(a).
\end{equation}
Taking the derivative with respect to the scale factor $a$, we obtain
\begin{equation}
\frac{d \ln H(a)}{da}
= \frac{1}{2(1+\alpha)} X^{-1}(a) \frac{dX(a)}{da}.
\end{equation}
Using Eq.~(\ref{eq:X}), we have
\begin{equation}
\frac{dX(a)}{da}
= -3(1+\alpha)\Omega_{m0} a^{-3(1+\alpha)-1}.
\end{equation}
Substituting this result, we obtain
\begin{equation}
\frac{d \ln H(a)}{da}
= -\frac{3}{2}\frac{\Omega_{m0} a^{-3(1+\alpha)-1}}{X(a)}.
\end{equation}
Finally, substituting into Eq.~(\ref{eq:q_a}), the deceleration parameter reads
\begin{equation}
q(a) = -1 + \frac{3}{2}\frac{\Omega_{m0} a^{-3(1+\alpha)}}{X(a)}.
\end{equation}
Alternatively, one can write
\begin{equation}\label{eq:qa}
q(a) = -1 + \frac{3}{2}\,
\frac{\Omega_{m0}\, a^{-3(1+\alpha)}}
{1 - \Omega_{m0} + \Omega_{m0} a^{-3(1+\alpha)}}.
\end{equation}

\subsection{Derivation of the total effective EoS parameter of the unified cosmic fluid}
\subsubsection{Derivation from the continuity equation}

We considered a spatially flat, radiation-free Universe ($\Omega_{k0} = \Omega_{r0} = 0$).
The continuity equation for the total cosmic fluid reads
\begin{equation}
\dot{\rho}_{tot} + 3H(\rho_{tot} + p_{tot}) = 0,
\end{equation}
where $\rho_{tot}$ and $p_{tot}$ denote the total energy density and pressure, respectively.
Defining the total effective EoS parameter as
\begin{equation}
w_{tot} = \frac{p_{tot}}{\rho_{tot}},
\end{equation}
the continuity equation becomes
\begin{equation}
\dot{\rho}_{tot} + 3H \rho_{tot} (1 + w_{tot}) = 0.
\end{equation}
Dividing by $\rho_{tot}$ yields
\begin{equation}
\frac{\dot{\rho}_{tot}}{\rho_{tot}} = -3H(1 + w_{tot}).
\end{equation}
Using
\begin{equation}
\frac{d \ln \rho_{tot}}{dt} = \frac{\dot{\rho}_{tot}}{\rho_{tot}}, 
\qquad
\frac{d \ln a}{dt} = H(a),
\end{equation}
we obtain
\begin{equation}
\frac{d \ln \rho_{tot}}{d \ln a} = -3(1 + w_{tot}).
\end{equation}
Solving for $w_{tot}$ gives
\begin{equation}
w_{tot}(a) = -1 - \frac{1}{3} \frac{d \ln \rho_{tot}}{d \ln a}.
\end{equation}

Since the total energy density scales as $\rho_{tot} \propto H^2(a) \propto E^2(a)$, following the first Friedmann equation $\rho_{tot} = 3H^2(a)$ (in units where $8\pi G = 1$), where the dimensionless Hubble parameter $E(a)$ is defined in Eq.~(\ref{eq:E}) under the assumptions $\Omega_{r0}=0$ and $\Omega_{k0}=0$, the above expression can be rewritten as
\begin{equation}\label{eq:wtot}
w_{tot}(a)=-1-\frac{1}{3}\frac{d\ln E^2(a)}{d\ln a}.
\end{equation}
Using Eq.~(\ref{eq:E}), we have
\begin{equation}
E^2(a)=X^{\frac{1}{1+\alpha}}(a),
\end{equation}
which implies 
\begin{equation}
\ln E^2(a)=\frac{1}{1+\alpha}\ln X(a).
\end{equation}
Taking the derivative with respect to $\ln a$, we get
\begin{equation}\label{eq:dlnEsq}
\frac{d\ln E^2(a)}{d\ln a}=\frac{1}{1+\alpha}\frac{1}{X(a)}\frac{dX(a)}{d\ln a}.
\end{equation}
The derivative of $X(a)$ is
\begin{equation}
\frac{dX(a)}{d\ln a}=-3(1+\alpha)\Omega_{m0}a^{-3(1+\alpha)}.
\end{equation}
Substituting this into Eq.~(\ref{eq:dlnEsq}), we obtain the following
\begin{equation}\label{eq:dlnEsqdlna}
\frac{d\ln E^2(a)}{d\ln a} =
-\frac{3\Omega_{m0}a^{-3(1+\alpha)}}{X(a)}.
\end{equation}
Finally, substituting into Eq.~(\ref{eq:wtot}), we derive the total effective EoS parameter
\begin{equation}\label{eq:wtot10}
w_{tot}(a)=
-1+
\frac{\Omega_{m0}a^{-3(1+\alpha)}}{X(a)}.
\end{equation}
Equivalently, this can be written as
\begin{equation}\label{eq:w_tot}
w_{tot}(a)=
-1+\frac{\Omega_{m0}a^{-3(1+\alpha)}}
{1-\Omega_{m0}+\Omega_{m0}a^{-3(1+\alpha)}}.
\end{equation}
 
\subsubsection{Consistency check from the second Friedmann equation}

In a spatially flat, radiation-free Universe ($\Omega_{k0} = \Omega_{r0} = 0$), and using units where $\kappa = 1$, the second Friedmann (acceleration) equation reads
\begin{equation} \label{eq:Ein}
\frac{\ddot{a}}{a} = -\frac{1}{6}\left(1 + 3 w_{tot}\right)\rho_{tot}.
\end{equation}
Inserting Eq.~(\ref{eq:q1}) into Eq.~(\ref{eq:Ein}) gives the following
\begin{equation}
q(a) = \frac{1}{6}\left(1 + 3 w_{tot}(a)\right)\frac{\rho_{tot}}{H^2(a)}
\end{equation}
or
\begin{equation}
q(a) = \frac{1}{2}\left(1 + 3 w_{tot}(a)\right)\Omega_{tot}(a).
\end{equation}
The  total density parameter of the fluid is defined as $\Omega_{tot}(a)=\rho_{tot}/\rho_{cr}$, with  the critical energy density $\rho_{cr} = 3H^{2}(a)$.
In a spatially flat Universe, the total energy density is equal to the critical density, implying $\Omega_{tot}(a) = 1$, which leads to the following
\begin{equation}
q(a) = \frac{1}{2}\left(1 + 3 w_{tot}(a)\right).
\end{equation}
Thus,
\begin{equation}\label{eq:wtot1}
w_{tot}(a) = \frac{2q(a) - 1}{3}.
\end{equation}
Inserting Eq.~(\ref{eq:qa}) for the deceleration parameter into Eq.~(\ref{eq:wtot1}) yields Eq.~(\ref{eq:w_tot})  for the total effective  EoS parameter of the cosmic fluid.

\section{Reconstruction of the DE EoS parameter}
\label{app:B}

We considered a spatially flat, radiation-free Universe ($\Omega_{k0} = \Omega_{r0} = 0$). The EoS parameter of the DE fluid is defined as
\begin{equation}\label{eq:w_de}
w_{DE} = \frac{p_{DE}}{\rho_{DE}}.
\end{equation}
Inserting $\rho_{tot} = \rho_{DE} + \rho_{m}$ into Eq.~(\ref{eq:Ein})  and using the definition of Eq.~(\ref{eq:w_de}) gives
\begin{equation}\label{eq:Ein1}
\frac{\ddot{a}}{a} = -\frac{1}{6}\bigg[(1 + 3w_{DE}(a))\rho_{DE} + \rho_{m}\bigg].
\end{equation}
In terms of density parameters, Eq.~(\ref{eq:Ein1}) takes the form
\begin{equation}\label{eq:Ein2}
\frac{\ddot{a}}{a} = -\frac{1}{2}\bigg[(1 + 3w_{DE}(a))\Omega_{DE}(a) + 
\Omega_{m}(a)\bigg]H^2(a).
\end{equation}
Using the relation $\Omega_{DE}(a) + \Omega_{m}(a) = 1$ in a spatially flat Universe, Eq.~(\ref{eq:Ein2}) reduces to
\begin{equation}
\frac{\ddot{a}}{a} = -\frac{1}{2}\bigg[1 + 3w_{DE}(a)\Omega_{DE}(a)\bigg]H^2(a).
\end{equation}
Hence, the deceleration parameter can be written as
\begin{equation}
q(a) = \frac{1}{2}(1 + 3w_{DE}(a)\Omega_{DE}(a)).
\end{equation}
Thus, the DE EoS parameter can be expressed in terms of the deceleration parameter as
\begin{equation}\label{eq:w_de0}
w_{DE}(a) = \frac{2q(a) - 1}{3\Omega_{DE}(a)}.
\end{equation}
Equivalently, using $\Omega_{DE}(a) = 1 - \Omega_{m}(a)$ in a spatially flat Universe, Eq.~(\ref{eq:w_de0}) can be rewritten as
\begin{equation} \label{eq:w_de1}
w_{DE}(a) = \frac{2 q(a) - 1}{3 \big(1 - \Omega_m(a)\big)}, 
\quad \text{with} \quad 
\Omega_m(a) = \frac{\Omega_{m0} a^{-3(1+\alpha)}}{E^2(a)}.
\end{equation}
Substituting Eq.~(\ref{eq:qa}) into Eq.~(\ref{eq:w_de1}), we obtain the explicit form of the DE EoS parameter in the $\Lambda(t)$CDM model
\begin{equation}\label{eq:w_DE} 
w_{DE}(a)=-\frac{1-\Omega_{ m0}a^{-3(1+\alpha)}X^{-1}(a)}{1-\Omega_{m0}a^{-3(1+\alpha)}X^{-\frac{1}{1+\alpha}}(a)},~{\rm with}~ \alpha\neq-1.
\end{equation}
For $\alpha = 0$, this expression reduces to the limit $w_{DE}(a) = -1$ of the standard $\Lambda$CDM model.

It is important to note that $w_{DE}(a)$ is reconstructed as $w_{DE}=p_{DE}/\rho_{DE}$. In this approach, $w_{DE}(a)$ can exhibit divergences at finite scale factor values when the  DE density approaches zero. This behaviour reflects a limitation of the reconstruction procedure rather than a physical singularity. In such regimes, $w_{DE}(a)$ should be interpreted as an effective diagnostic quantity rather than a fundamental physical parameter.

\end{document}